\documentclass[12pt]{article}

\usepackage{graphicx}
\usepackage{float}

\def\gtwid{\mathrel{\raise.3ex\hbox{$>$\kern-.75em\lower1ex\hbox{$\sim$}}}}
\def\ltwid{\mathrel{\raise.3ex\hbox{$<$\kern-.75em\lower1ex\hbox{$\sim$}}}}
\def\square{\kern1pt\vbox{\hrule height 1.2pt\hbox{\vrule width 1.2pt\hskip 3pt
   \vbox{\vskip 6pt}\hskip 3pt\vrule width 0.6pt}\hrule height 0.6pt}\kern1pt}

\begin{document}

\begin{titlepage}

\begin{flushright}
UFIFT-QG-20-01
\end{flushright}

\vskip 0.2cm

\begin{center}
{\bf Bose-Fermi Cancellation of Cosmological Coleman-Weinberg Potentials}
\end{center}

\vskip 0.2cm

\begin{center}
S. P. Miao$^{1*}$, L. Tan$^{2\star}$ and R. P. Woodard$^{2\dagger}$
\end{center}

\begin{center}
\it{$^{1}$ Department of Physics, National Cheng Kung University, \\
No. 1 University Road, Tainan City 70101, TAIWAN}
\end{center}

\begin{center}
\it{$^{2}$ Department of Physics, University of Florida,\\
Gainesville, FL 32611, UNITED STATES}
\end{center}

\vspace{0.2cm}

\begin{center}
ABSTRACT
\end{center}
Cosmological Coleman-Weinberg potentials are induced when normal matter is 
coupled to the inflaton. It has long been known that the corrections from
bosonic fields are positive whereas those from fermionic fields are negative.
In flat space both take the form $\pm \varphi^4 \ln(\varphi)$, and they can 
be made to cancel by appropriately choosing the coupling constants. In an
expanding universe the bosonic and fermionic results no longer take the same
form, although their large field limits do. We choose the coupling constants
so that the large field limits cancel, and then follow the deviations which
result as inflation progresses. Although the result is not satisfactory we
discuss how adding scalars with various conformal couplings likely solves
the problem.

\begin{flushleft}
\end{flushleft}

\vspace{0.2cm}

\begin{flushleft}
$^{*}$ e-mail: spmiao5@mail.ncku.edu.tw \\
$^{\star}$ email:  ltan@ufl.edu \\
$^{\dagger}$ e-mail: woodard@phys.ufl.edu
\end{flushleft}

\end{titlepage}

\section{\large Introduction}

The case for an early phase of accelerated expansion is powerfully
supported by cosmological data \cite{Geshnizjani:2011dk,Kinney:2012zbd} 
but there is not yet any compelling indication of what caused it. The 
latest data \cite{Akrami:2018odb} are consistent with the simplest model 
based on the potential of a minimally coupled scalar,
\begin{equation}
\mathcal{L} = \frac{R \sqrt{-g}}{16 \pi G} - \frac12 \partial_{\mu}
\varphi \partial_{\nu} \varphi g^{\mu\nu} \sqrt{-g} - V(\varphi) 
\sqrt{-g} \; . \label{scalardriven}
\end{equation}
However, this class of models comes with a high burden of fine tuning
in order to make inflation start, to make it last long enough, to 
generate primordial perturbations of the observed strength, and to
avoid losing predictability through the formation of a multiverse
\cite{Ijjas:2013vea}. No one disputes these facts, but the wildly
differing interpretations of them \cite{Guth:2013sya,Linde:2014nna,
rebuttal} has been termed an ``inflationary schism'' \cite{Ijjas:2014nta}.

We are concerned with another sort of fine-tuning problem that derives from
coupling $\varphi$ to conventional particles to give efficient re-heating.
It has long been known that the 0-point motion of the coupled particles 
induces Coleman-Weinberg \cite{Coleman:1973jx} corrections to $V(\varphi)$.
Because these corrections are not Planck-suppressed they are unacceptably
large \cite{Green:2007gs}. Further, they cannot be completely absorbed into
local modifications of the Lagrangian (\ref{scalardriven}) because they 
involve complicated functions of the dimensionless ratio of $\varphi$ to the
Hubble parameter \cite{Miao:2015oba} on the de Sitter background to which
computations have so far been limited. 

A approximation for the effective potential from a scalar on a general 
inflationary background \cite{Kyriazis:2019xgj} indicates the rough validity 
of assuming that the constant de Sitter Hubble parameter of existing 
computations becomes the time-dependent Hubble parameter for a general 
cosmological background. Because the Hubble parameter is not even a local 
functional of the metric generally, there are only two sorts of local 
Lagrangians that can be used to cancel cosmological Coleman-Weinberg 
potentials:\footnote{The Appendix discusses using proxies for the Hubble
parameter.}
\begin{enumerate}
\item{Subtract a function of only the inflaton \cite{Liao:2018sci}; or}
\item{Subtract a function of the inflaton and the Ricci scalar \cite{Miao:2019bnq}.}
\end{enumerate}
Neither subtraction leads to acceptable results, and locality, invariance 
and stability preclude any more general subtraction \cite{Woodard:2006nt}.

Because subtractions cause so many problems \cite{Liao:2018sci,Miao:2019bnq}
we wish here to explore the viability of canceling the cosmological 
Coleman-Weinberg potentials induced by coupling $\varphi$ to a boson with 
those derived from coupling to a fermion. In flat space such a cancellation
would be exact because bosonic contributions go like $+\varphi^4 \ln(\varphi)$
and fermionic contributions go like $-\varphi^4 \ln(\varphi)$ 
\cite{Coleman:1973jx}. However, the existing results on de Sitter background
are quite different for bosons and fermions, although they of course approach 
the usual flat space forms in the large $\varphi$ (small Hubble parameter) 
regime \cite{Miao:2015oba}. We shall accordingly (in section 2) choose the 
coupling constants to make the cancellation exact in the flat space limit, 
and then (in section 3) study the effect on inflation of the incomplete 
cancellation for nonzero Hubble parameter. Our conclusions comprise section 4.

\section{\large Cosmological Coleman-Weinberg Potentials}

The purpose of this section is to present the cosmological Coleman-Weinberg
potentials whose effect on inflation is the subject of this paper. We begin
by reviewing the contributions from a Yukawa-coupled fermion and from a vector
boson. We then give the large field limits of each contribution and derive the
relation between the Yukawa coupling constant $h$ and the vector boson coupling
$q$ that makes the large field limits cancel. The section closes by expressing 
everything in a convenient dimensionless form.  

\subsection{\normalsize Effective potentials from fermions and vector bosons}

In order to include both fermionic and vector boson couplings we change the 
classical model (\ref{scalardriven}) from a real to a complex scalar inflaton 
$\varphi$,
\begin{eqnarray}
\lefteqn{\mathcal{L} = \frac{R \sqrt{-g}}{16 \pi G} \!-\! \frac14 F_{\rho\sigma} 
F_{\mu\nu} g^{\rho\mu} g^{\sigma\nu} \sqrt{-g} + \overline{\psi} e^{\mu}_{~b} 
\gamma^b \Bigl[i\partial_{\mu} \!-\! S_{\mu} \Bigl] \psi \sqrt{-g} \!-\! h \vert 
\varphi\vert \overline{\psi} \psi \sqrt{-g} } \nonumber \\
& & \hspace{-0.7cm} - \Bigl(\partial_{\mu} \!-\! i q A_{\mu} \Bigr) \varphi^*
\Bigl( \partial_{\nu} \!+\! i q A_{\nu}\Bigr) \varphi g^{\mu\nu} \sqrt{-g} 
- \Bigl[ V(\vert \varphi\vert) \!+\! \delta \xi \vert \varphi \vert^2 R \!+\! 
\frac{\delta \lambda}{4} \vert \varphi \vert^4 \Bigr] \sqrt{-g} . \quad 
\label{totalL}
\end{eqnarray}
Here $h$ and $q$ are the Yukawa and electromagnetic coupling constants, 
respectively, whereas $\delta \xi$ and $\delta \lambda$ are counterterms. 
The vector boson field is $A_{\mu}$ and $F_{\mu\nu} \equiv \partial_{\mu} 
A_{\nu} - \partial_{\nu} A_{\mu}$ is its field strength tensor. The Dirac 
fermion field is $\psi$ ($\overline{\psi} \equiv \psi^{\dagger} \gamma^0$), 
with gamma matrices $\gamma^a$ ($\{\gamma^b , \gamma^c\} = -2 \eta^{bc}$), 
vierbein $e_{\mu b}$ ($g_{\mu\nu} = e_{\mu b} e_{\nu c} \eta^{bc}$), and
spin connection matrices $S_{\mu} \equiv \frac{i}8 [\gamma^b , \gamma^c] 
e^{\nu}_{~c} ( e_{\nu d,\mu} - \Gamma^{\rho}_{~\mu\nu} e_{\rho d})$.

The geometry is,
\begin{equation}
g_{\mu\nu} dx^{\mu} dx^{\nu} = -dt^2 + a^2(t) d\vec{x} \!\cdot\! d\vec{x}
\qquad , \qquad H(t) \equiv \frac{\dot{a}}{a} \; . \label{background}
\end{equation}
The fermionic and vector bosonic contributions to the inflaton effective
potential are so far only known for the de Sitter case of exactly constant 
$H$. The original computations \cite{Candelas:1975du,Allen:1983dg} used
various regularization and renormalization conventions. However, when
dimensional regularization (in spacetime dimension $D$ with renormalization
scale $\mu$) is employed with counterterms,
\begin{eqnarray}
\delta \xi \!\!\!& = &\!\!\! \frac{2 h^2 \mu^{D-4}}{(4 \pi)^{\frac{D}2}} 
\Biggl\{\frac{\Gamma(1 \!-\! \frac{D}2)}{D (D \!-\! 1)} + 
\frac{1 \!-\! \gamma}{6}\Biggr\} + \frac{ q^2 \mu^{D-4}}{(4 \pi)^{\frac{D}2}} 
\Biggr\{ \frac1{4 \!-\! D} + \frac{\gamma}2\Biggr\} \; , \label{deltaxi} \\
\delta \lambda \!\!\!& = &\!\!\! \frac{4 h^4 \mu^{D-4}}{(4 \pi)^{\frac{D}2}} 
\Biggl\{\Gamma\Bigl(1 \!-\! \frac{D}2\Bigr) \!+\! 2 \zeta(3) \!-\! 2 
\gamma\Biggr\} + \frac{12 q^4 \mu^{D-4}}{(4 \pi)^{\frac{D}2}} \Biggr\{ 
\frac2{4 \!-\! D} \!+\! \gamma \!-\! \frac83 \Biggr\} \; , \qquad 
\label{deltalambda}
\end{eqnarray}
($\gamma \simeq 0.577$ is Euler's constant and $\zeta(s)$ is the Riemann zeta 
function) the effective potential takes the form \cite{Miao:2006pn,
Prokopec:2007ak,Miao:2015oba,Miao:2019bnq},
\begin{eqnarray}
\lefteqn{V_{\rm eff} = -\frac{H^4}{8 \pi^2} \Biggl\{ f(h^2 z) + \Bigl[h^2 z 
\!+\! \frac{h^4 z^2}{2} \Bigr] \ln\Bigl(\frac{H^2}{\mu^2}\Bigr) \Biggr\} }
\nonumber \\
& & \hspace{5cm} + \frac{3 H^4}{8 \pi^2} \Biggl\{ b(q^2 z) + \Bigl[q^2 z 
\!+\! \frac{q^4 z^2}{2} \Bigr] \ln\Bigl(\frac{H^2}{\mu^2}\Bigr) \Biggr\} . 
\qquad \label{Veff}
\end{eqnarray}
Here $z \equiv \vert \varphi\vert^2/H^2$ and the functions $f(y)$ and $b(y)$
are expressed as integrals of the digamma function $\psi(x) \equiv \frac{d}{dx}
\ln[\Gamma(x)]$,
\begin{eqnarray}
f(y) & = & 2 \gamma y - \Bigl(\zeta(3) \!-\! \gamma\Bigr) y^2 + 2 \int_0^{\sqrt{y}} 
\!\!\!\! dx \, (x \!+\! x^3) \Bigl[\psi(1 \!+\! i x) + \psi(1 \!-\! i x)\Bigr] \; , 
\qquad \label{fdef} \\
b(y) & = & \Bigl(2 \gamma \!-\! 1\Bigr) y - \Bigl( \frac32 \!-\! \gamma\Bigr) y^2
\nonumber \\
& & \hspace{1.7cm} + \int_0^{y} \!\! dx \, (1 \!+\! x) \Biggl[ \psi\Bigl(\frac32 \!+\!
\frac{\sqrt{1 \!-\! 8 x}}{2} \Bigr) + \psi\Bigl(\frac32 \!-\! 
\frac{\sqrt{1 \!-\! 8x}}{2} \Bigr) \Biggr] \; . \qquad \label{bdef}
\end{eqnarray}

\subsection{\normalsize Canceling the large field limits}

The functional forms of the negative fermionic contribution to (\ref{Veff}) and the
positive bosonic contribution seem very different but they have the same large field
forms. This follows when one does the integrals of (\ref{fdef}) and (\ref{bdef})
term-by-term, using the large argument expansion of the digamma function,
\begin{equation}
\psi(x) = \ln(x) - \frac1{2x} - \frac1{12 x^2} + \frac1{120 x^4} - \frac1{256 x^6}
+ O\Bigl( \frac1{x^8}\Bigr) \; .
\end{equation}
The resulting expansions are \cite{Miao:2015oba},
\begin{eqnarray}
f(y) & = & \frac12 y^2 \ln(y \!+\! 1) - \Bigl( \zeta(3) \!+\! \frac14 \!-\! 
\gamma\Bigr) y^2 + y \ln(y \!+\! 1) \nonumber \\
& & \hspace{4cm} - \Bigl( \frac43 \!-\! 2 \gamma\Bigr) y + \frac{11}{60} \ln(y \!+\!1)
+ O(1) \; , \qquad \label{fexp} \\
b(y) & = & \frac12 y^2 \ln(y \!+\! 1) - \Bigl( \frac74 \!-\! \frac12 \ln(2) \!-\! 
\gamma\Bigr) y^2 + y \ln(y \!+\! 1) \nonumber \\
& & \hspace{3cm} - \Bigl( \frac{13}{6} \!-\! \ln(2) \!-\! 2 \gamma\Bigr) y + 
\frac{19}{60} \ln(y \!+\!1) + O(1) \; . \qquad \label{bexp}
\end{eqnarray}
For large $\vert \varphi\vert$ the leading fermionic and bosonic contributions to 
$V_{\rm eff}$ go like $\mp \vert\varphi\vert^4 \ln(\vert\varphi\vert)$. Setting $h =
3^{\frac14} q$ causes these terms to cancel, leaving a slightly negative remainder,
\begin{equation}
V_{\rm eff} \Bigl\vert_{h^2 = \sqrt{3} \, q^2} \longrightarrow -\frac{3 q^4 \vert
\varphi\vert^4}{8 \pi^2} \Biggl\{ \frac32 - \zeta(3) - \frac14 \ln(\frac{4}{3}\Bigr)
\Biggr\} + O\Biggl( q^2 H^2 \vert\varphi\vert^2 \ln(\vert \varphi\vert)\Biggr) \; .
\label{Vlead}
\end{equation}
The number inside the curly brackets of (\ref{Vlead}) is $K \equiv \frac32 - \zeta(3) - 
\frac14 \ln(\frac43) \simeq 0.23$.

As can already be seen from the negative sign of (\ref{Vlead}), the cancellation
scheme we have devised will only give a metastable model, that must decay to a Big
Rip singularity for a sufficiently large value of $q$. In Section 4 we discuss the
prospects for avoiding this problem by involving scalars with various conformal
couplings. We selected (\ref{totalL}) for this initial study because fermions and 
vector gauge bosons certainly exist in the Standard Model, whereas there is only
one known scalar, and its conformal coupling is a matter of conjecture.

Any sort of relation between coupling constants such as $h = 3^{\frac14} q$ is liable
to criticism as a fine tuning unless it follows from some symmetry. The obvious 
candidate here would be supersymmetry, but our Lagrangian (\ref{totalL}) is not 
supersymmetric for any values of the coupling constants $h$ and $q$, and we have
been unable to identify any symmetry which justifies the relation $h = 3^{\frac14} q$.
In view of the fact that this particular relation does not lead to a viable model of 
inflation we will content ourselves here with simply noting the importance, for future 
models that may be viable, of identifying some symmetry to protect against fine tuning. 

Higher loop corrections are also an issue. There are not now any higher loop results 
for cosmological Coleman-Weinberg potentials, even on de Sitter background. However,
flat space results are so large that even two and three loop corrections could be 
problematic for scalar-driven inflation. Therefore, any viable cancellation scheme 
must extend beyond one loop order. A related issue is renormalization group flows
for the various couplings, which represent a way of summing up large logarithms from
perturbative loop corrections. Inflation takes place at an indeterminate but likely 
very high energy scale, and that scale changes markedly over the course of inflation.
Hence any viable cancellation scheme must be stable under likely renormalization group 
flows. 
 
\subsection{\normalsize Dimensionless formulation}

\begin{figure}[H]
\includegraphics[width=6cm,height=6cm]{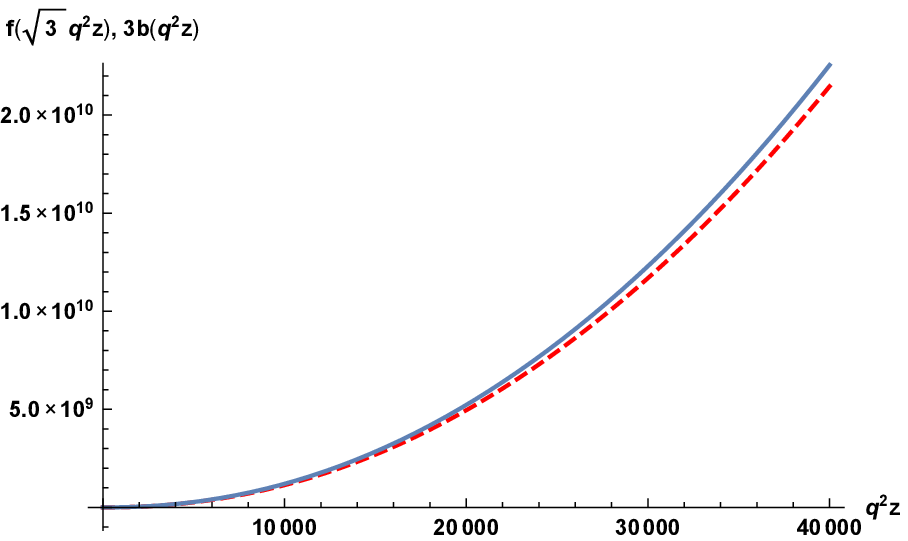}
\hspace{0.3cm}
\includegraphics[width=6cm,height=6cm]{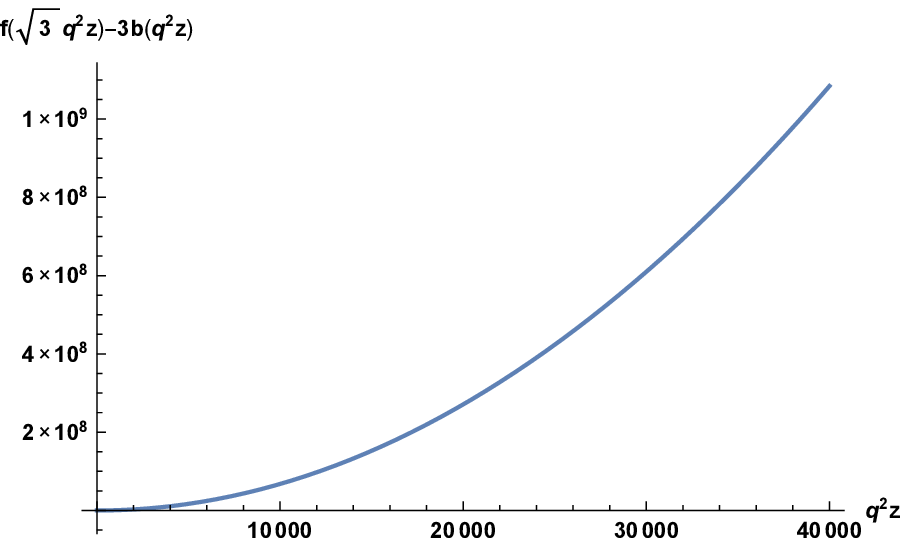}
\caption{The left hand figure shows the fermionic function $f(\sqrt{3} q^2 z)$ 
(in solid blue) which was defined in expression (\ref{fdef}), with the bosonic
function $3 b(q^2 z)$ (in dashed red) which was defined in expression (\ref{bdef}). 
The right hand figure shows the combination $f(\sqrt{3} q^2 z) - 3 b(q^2 z)$ that
appears in the effective potential (\ref{Ueff}). The coupling constant is $q^2 = 
5.0 \times 10^{-7}$.} 
\label{quantum}
\end{figure}

It is desirable to change the independent variable from co-moving time $t$ to
the dimensionless number of e-foldings from the start of inflation $n \equiv 
\ln[a(t)/a(t_i)]$. This carries derivatives to,
\begin{equation}
\frac{d}{dt} = H \frac{d}{d n} \qquad , \qquad \frac{d^2}{dt^2} = H^2 \Bigl[
\frac{d^2}{dn^2} - \epsilon \frac{d}{dn}\Bigr] \; , \label{dimless1}
\end{equation}
where $\epsilon(n) = -H'/H$ is the first slow roll parameter. We also extract a 
factor of $\sqrt{8 \pi G}$ from the inflaton, the Hubble parameter and the 
renormalization scale,
\begin{equation}
\phi(n) \equiv \sqrt{8 \pi G} \, \vert\varphi(t)\vert \quad , \quad \chi(n) \equiv 
\sqrt{8\pi G} \, H(t) \quad , \quad s \equiv \sqrt{8\pi G} \, \mu \; . 
\label{dimless2}
\end{equation}
And we extract a factor of $(8 \pi G)^2$ from the classical potential $V(\varphi)$
and the cosmological Coleman-Weinberg potential $V_{\rm eff}(\varphi,H)$,
\begin{equation}
U(\phi) \equiv (8 \pi G)^2 V(\vert\varphi\vert) \qquad , 
\qquad U_{\rm eff}(\phi,\chi) \equiv (8\pi G)^2 V_{\rm eff}(\vert\varphi\vert,H) 
\; . \label{dimless3}
\end{equation} 
With these definitions the dimensionless cosmological Coleman-Weinberg potential
from fermions (with $h^2 = \sqrt{3} q^2$) and vector bosons is,
\begin{equation}
U_{\rm eff}(\phi,\chi) = -\frac{\chi^4}{8\pi^2} \Biggl\{ f\Bigl(\sqrt{3} \, 
q^2 z\Bigr) - 3 b\Bigl(q^2 z\Bigr) - \Bigl(3 \!-\! \sqrt{3}\,\Bigr) q^2 z 
\ln\Bigl( \frac{\chi^2}{s^2} \Bigr) \Biggr\} \; , \label{Ueff}
\end{equation}
where $z = \phi^2/\chi^2$. Figure~\ref{quantum} shows the individual fermionic and
bosonic contributions to $U_{\rm eff}$, as well as their sum.

\section{\large Evolution during Inflation}

In this section we quantify how cosmological Coleman-Weinberg potentials 
modify classical inflation for the simple quadratic potential. The section
begins by comparing the exact numerical evolution of the classical model 
with its slow roll approximation. Then the effect of the quantum correction 
is studied for various values of the coupling constant $q$.

\subsection{\normalsize Evolution of the Classical Model}

The scalar evolution equation is,
\begin{equation}
\phi'' + (3 \!-\! \epsilon) \phi' + \frac{U'(\phi)}{2 \chi^2} = 0 \; .
\label{classicalscalar}
\end{equation}
The dimensionless Hubble parameter and the first slow roll parameter are
computed from $\phi(n)$ through the relations,
\begin{equation}
\chi^2(n) = \frac{U(\phi(n))}{3 \!-\! {\phi'}^2(n)} \qquad , \qquad
\epsilon(n) = {\phi'}^2(n) \; . \label{classicalgeometry}
\end{equation}
The required initial value data is obviously $\phi_0 \equiv \phi(0)$ and
$\phi'_0 \equiv \phi'(0)$.

\begin{figure}[H]
\includegraphics[width=4.5cm,height=4.5cm]{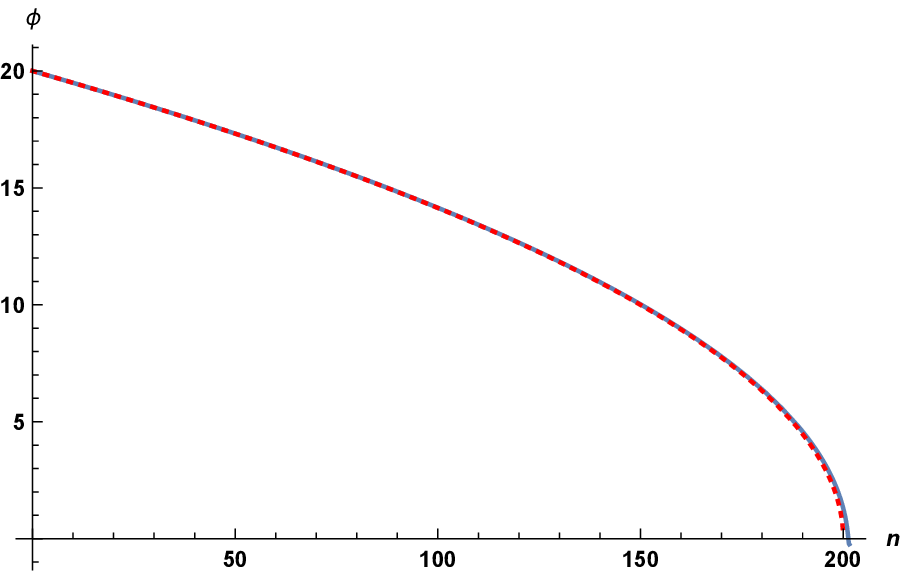}
\hspace{-0.3cm}
\includegraphics[width=4.5cm,height=4.5cm]{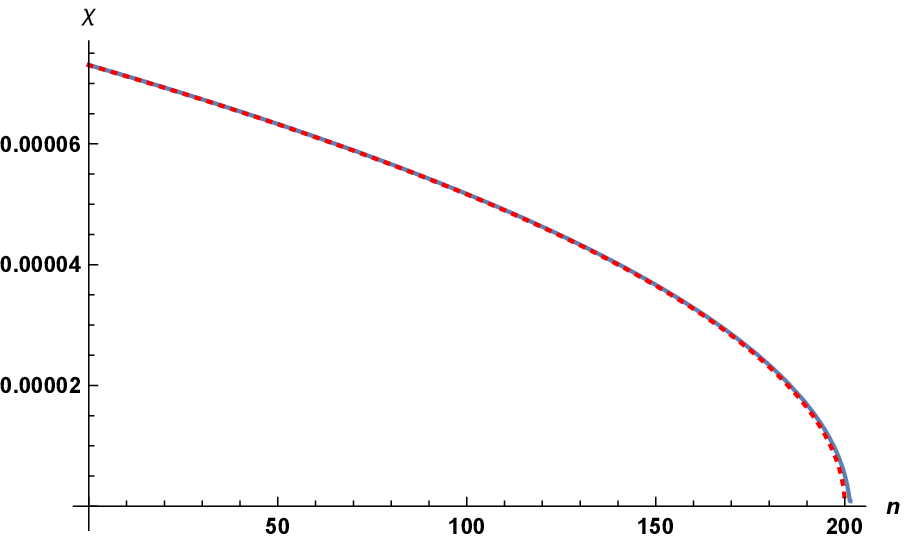}
\hspace{-0.3cm}
\includegraphics[width=4.5cm,height=4.5cm]{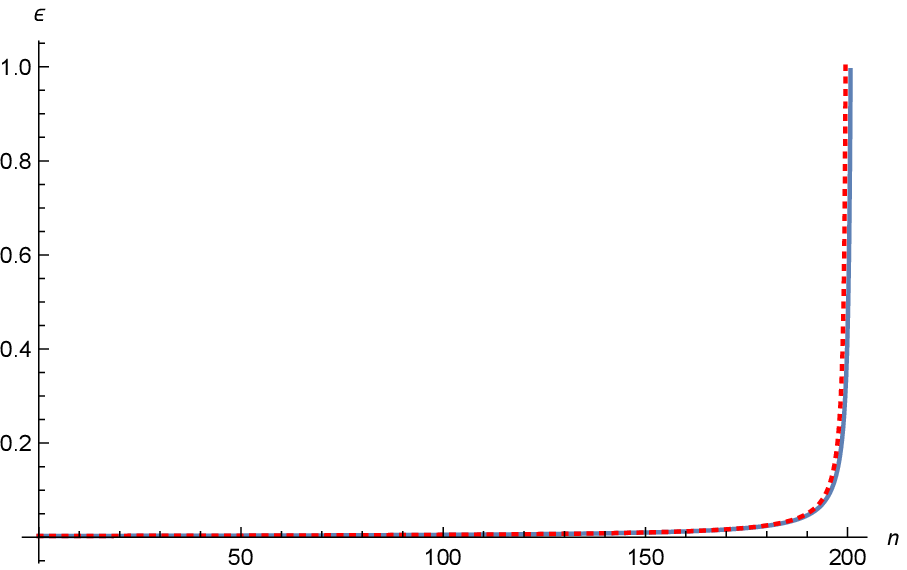}
\caption{These figures show the classical evolutions (solid blue) 
versus the slow roll approximations (dashed red) with $k^2 = 4 
\times 10^{-11}$. Shown are the dimensionless scalar $\phi(n)$ 
(on the left), the dimensionless Hubble parameter $\chi(n)$ (in 
the middle), and the first slow roll parameter $\epsilon(n)$ 
(on the right).}
\label{classical}
\end{figure}

We choose the dimensionless classical potential to be $U(\phi) = k^2 \phi^2$,
with $k^2 = 4 \times 10^{-11}$. Making the slow roll approximation allows us
to solve equations (\ref{classicalscalar}-\ref{classicalgeometry}) in terms 
of just $\phi_0$,
\begin{equation}
\phi(n) \simeq \sqrt{\phi_0^2 \!-\! 2 n} \quad , \quad \chi(n) \simeq 
\frac{k}{\sqrt{3}} \sqrt{ \phi_0^2 \!-\! 2 n} \quad , \quad \epsilon(n) 
\simeq \frac{1}{\phi_0^2 \!-\! 2 n} \; . \label{slowroll}
\end{equation}
Choosing $\phi_0 = 20$ corresponds to about 200 e-foldings of inflation. 
Figure~\ref{classical} compares numerical evolution of 
(\ref{classicalscalar}-\ref{classicalgeometry}) with the slow roll predictions
(\ref{slowroll}) for initial value data,
\begin{equation}
\phi_0 = 20 \qquad , \qquad \phi'_0 = -\frac1{20} \; . \label{initial}
\end{equation}
Agreement is excellent.

Primordial cosmological perturbations furnish the principal observable for 
primordial inflation. In the leading slow roll approximation the scalar and
tensor power spectra which experience first horizon crossing $n$ e-foldings
after the beginning of inflation are,
\begin{equation}
\Delta^2_{\mathcal{R}}(n) \simeq \frac1{8 \pi^2} \!\times\! 
\frac{\chi^2(n)}{\epsilon(n)} \qquad , \qquad \Delta^2_{h}(n) \simeq 
\frac1{8 \pi^2} \!\times\! 16 \chi^2(n) \; . \label{power}
\end{equation}
This makes the scalar spectral index and the tensor-to-scalar ratio,
\begin{equation}
1 - n_s(n) \simeq 2 \epsilon(n) + \frac{\epsilon'(n)}{\epsilon(n)} \qquad ,
\qquad r(n) \simeq 16 \epsilon(n) \; . \label{index}
\end{equation}
From the slow roll results (\ref{slowroll}) $\epsilon' \simeq 2 \epsilon^2$.
Evaluating (\ref{index}) at $n = 150$ (which is about 50 e-foldings before 
the end of inflation) gives,
\begin{equation}
1 - n_s(150) \simeq 0.04 \qquad , \qquad r(150) \simeq 00.16 \; .
\label{results}
\end{equation}
The measured scalar spectral index $n_s = 0.0351 \pm 0.0042$ agrees well 
with (\ref{results}), but the $95\%$ confidence limit of $r < 0.056$ is
significantly discrepant \cite{Akrami:2018odb}. Of course this invalidates
the quadratic model but we will continue to employ it on account of its
simplicity, and the fact that the problem we find is even worse for the 
more realistic (flatter) potentials that are still consistent with current 
data.   

\subsection{\normalsize Evolution of the Quantum-Corrected Model}

When $U_{\rm eff}(\phi,\chi)$ is added to the Lagrangian the scalar
evolution equation becomes,
\begin{equation}
\phi'' + (3 \!-\! \epsilon) \phi' + \frac1{2 \chi^2} \Bigl[ 
\frac{\partial U}{\partial \phi} + \frac{\partial U_{\rm eff}}{\partial \phi}
\Bigr] = 0 \; . \label{quantumscalar}
\end{equation}
The two nontrivial Einstein equations are \cite{Liao:2018sci},
\begin{eqnarray}
3 \chi^2 & = & \chi^2 {\phi'}^2 + U + U_{\rm eff} - \chi
\frac{\partial U_{\rm eff}}{\partial \chi} \; , \label{quantumE1} \\
-(3 \!-\! 2\epsilon) \chi^2 & = & \chi^2 {\phi'}^2 - U - U_{\rm eff} +
\chi \frac{\partial U_{\rm eff}}{\partial \chi} + \frac13 \chi \frac{d}{dn}
\frac{\partial U_{\rm eff}}{\partial \chi} \; . \qquad \label{quantumE2}
\end{eqnarray}
We obtain the evolution equation for $\chi(n)$ from the sum of (\ref{quantumE1})
and (\ref{quantumE2}),
\begin{equation}
\chi' = -\Biggl[ \frac{\chi {\phi'}^2 + \frac16 \phi' 
\frac{\partial^2 U_{\rm eff}}{\partial \phi \partial \chi}}{1 + \frac16
\frac{\partial^2 U_{\rm eff}}{\partial \chi^2}} \Biggr] \; , \label{quantumchi}
\end{equation}
with the initial value $\chi(0)$ numerically determined from (\ref{quantumE1}).
Evolving $\chi(n)$, rather than inferring it from $\phi(n)$, might seem a major
departure from the classical system (\ref{classicalgeometry}). However, setting
$U_{\rm eff} = 0$ reduces the evolution equation (\ref{quantumchi}) for $\chi(n)$ 
to the same relation $\chi' \equiv -\chi \epsilon = -\chi {\phi'}^2$ 
(\ref{classicalgeometry}) that could have been used to evolve the classical 
system. This means that quantum corrections are merely perturbing classical
solutions, rather than introducing new degrees of freedom which would be 
problematic \cite{Simon:1990ic}.

Our classical potential is $U(\phi) = +k^2 \phi^2$, whereas one can see from
Figure~\ref{quantum} that the quantum correction $U_{\rm eff}(\phi,\chi)$ is
negative definite. From the quartic growth of the large field form (\ref{Vlead}) 
we see that the total potential $U(\phi) + U_{\rm eff}(\phi,\chi) \simeq k^2 \phi^2
- 3 K q^4 \phi^4/8\pi^2$ is unbounded below at large $\phi$. We therefore expect 
that evolution depends on whether the scalar is initially driven inward to $\phi 
\rightarrow 0$ or outward to $\phi \rightarrow \infty$. With fixed initial 
conditions (\ref{initial}) this is controlled by the coupling constant $q^2$: 
for sufficiently small $q^2$ the scalar rolls in towards $\phi \rightarrow 0$, 
and there can be a graceful exit from inflation, but for larger values of $q^2$ 
the scalar rolls outwards towards $\phi \rightarrow \infty$ and the universe ends 
in a Big Rip singularity. Using the large field limiting form (\ref{Vlead}) we
estimate the threshold value of $q^2$ to be about,
\begin{equation}
U(\phi) + U_{\rm eff}(\phi,\chi) \longrightarrow k^2 \phi^2 - 
\frac{3 K q^4 \phi^4}{8 \pi^2} \Longrightarrow q^2_{*}(\phi_0) \simeq 
\frac{2 \pi k}{\sqrt{3 K} \, \phi_0} \simeq 2.4 \times 10^{-6} \; . \label{qcrit}
\end{equation}

\begin{figure}[H]
\includegraphics[width=4.5cm,height=4.5cm]{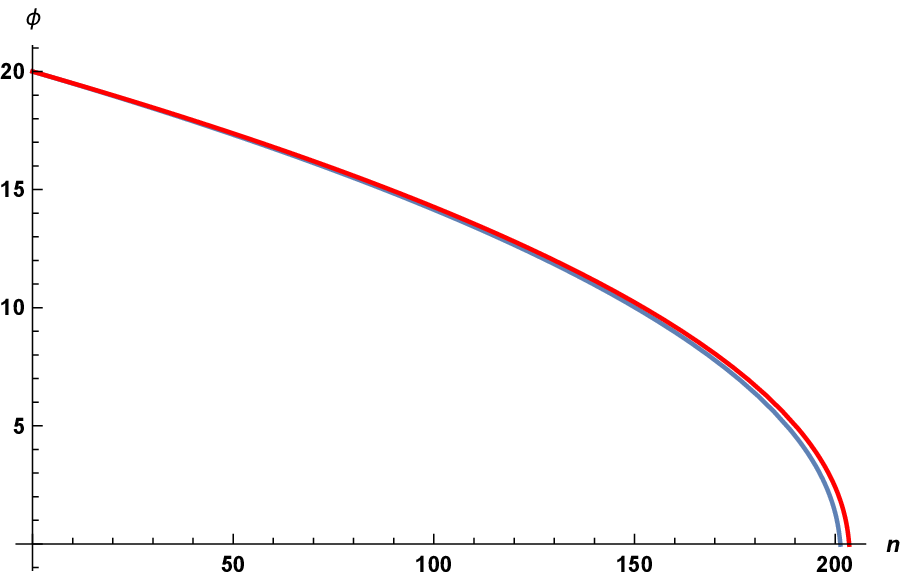}
\hspace{-0.3cm}
\includegraphics[width=4.5cm,height=4.5cm]{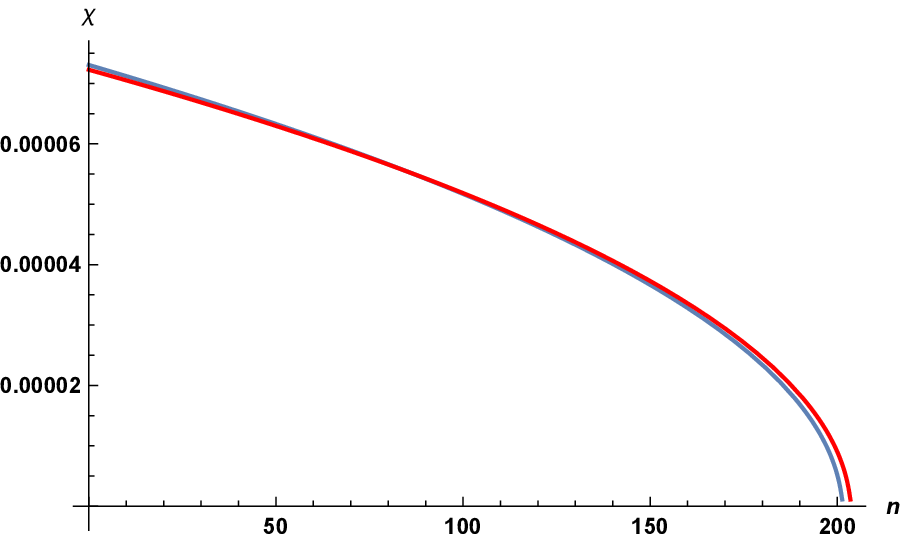}
\hspace{-0.3cm}
\includegraphics[width=4.5cm,height=4.5cm]{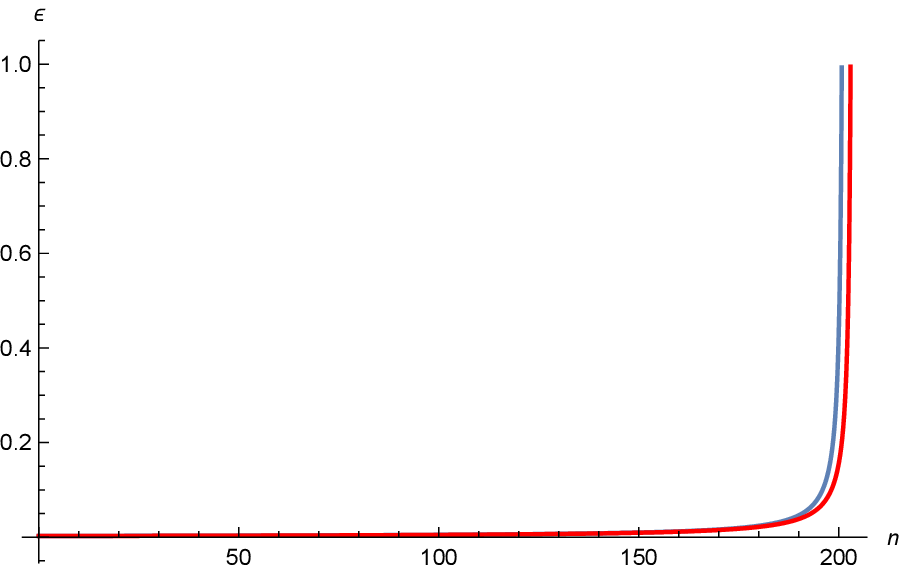}
\caption{These figures show the classical (blue) and quantum (red)
evolutions of $\phi(n)$, $\chi(n)$ and $\epsilon(n)$ for $q^2 = 5 \times 10^{-7}
< q^2_{*}(\phi_0)$.}
\label{smallq}
\end{figure}

Explicit numerical evolution confirms these expectations. Figure~\ref{smallq}
compares evolution for the classical and quantum systems with the small coupling
of $q^2 = 5.0 \times 10^{-7} < q^2_{*}(\phi_0)$.
\noindent 
Because this particular model of classical inflation has $\chi(n) \simeq k 
\phi(n)/\sqrt{3}$, the parameter $q^2$ times $z \equiv \phi^2/\chi^2$ on which 
the functions $f(\sqrt{3} q^2 y)$ and $3 b(q^2 y)$ depend is always in the large 
field regime,
\begin{equation}
q^2 z \simeq \frac{3 q^2}{k^2} \simeq 38,000 \; . \label{parameter}
\end{equation}
Hence the quartic large field limit (\ref{Vlead}) of the quantum correction
is valid throughout inflation, which makes the quantum correction less and less
important as the scalar rolls towards zero.

\begin{figure}[H]
\includegraphics[width=4.5cm,height=4.5cm]{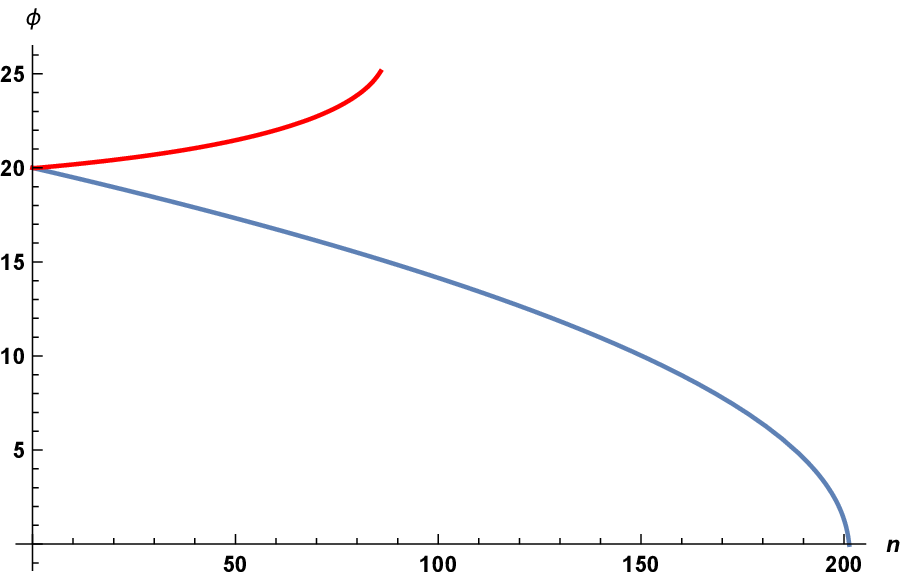}
\hspace{-0.3cm}
\includegraphics[width=4.5cm,height=4.5cm]{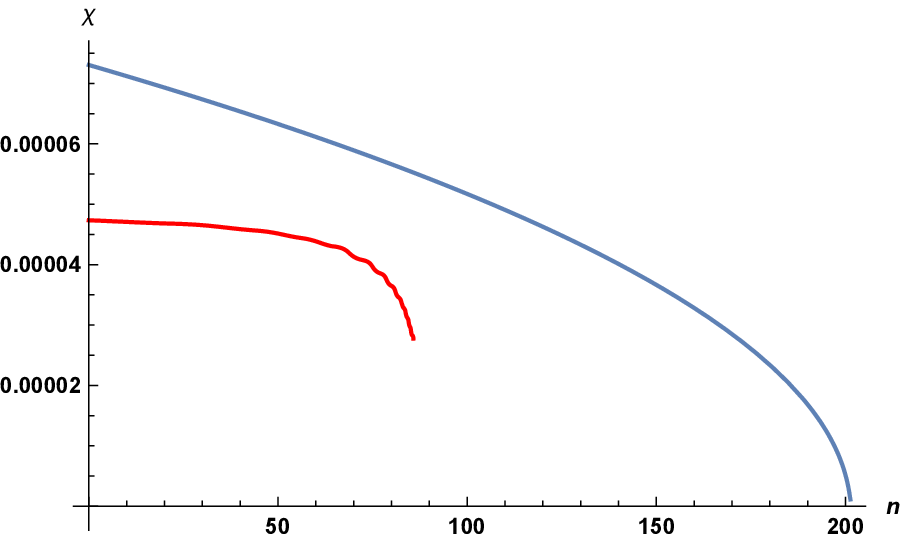}
\hspace{-0.3cm}
\includegraphics[width=4.5cm,height=4.5cm]{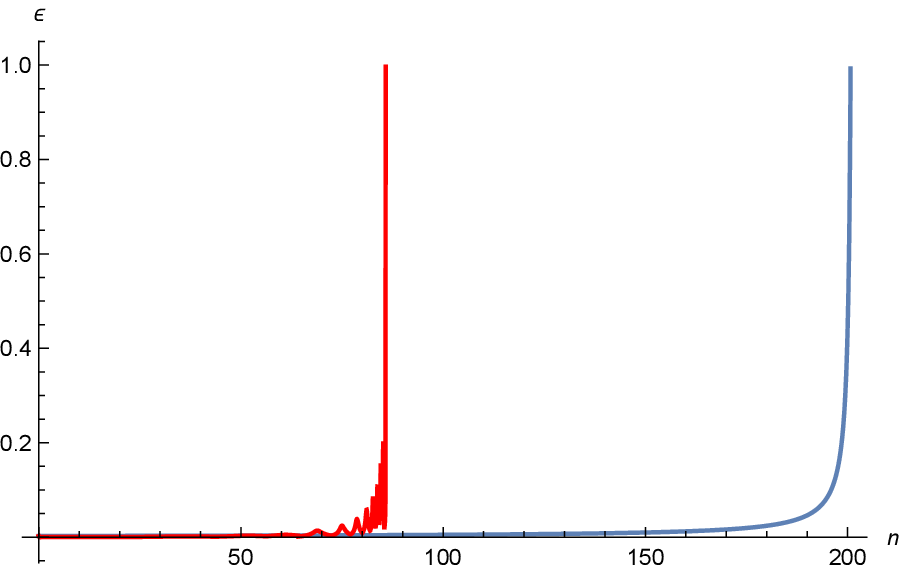}
\caption{These figures show the classical (blue) and quantum (red)
evolutions of $\phi(n)$, $\chi(n)$ and $\epsilon(n)$ for $q^2 = 2.6 \times 10^{-6} 
> q^2_{*}(\phi_0)$.}
\label{largeq}
\end{figure}
\noindent On the other hand, the slightly larger value of $q^2 = 2.6 \times 10^{-6} > 
q^2_{*}(\phi_0)$ leads to the disastrous evolution shown in Figure~\ref{largeq}.
The fact that $\chi(n)$ is driven to zero, while $\phi(n)$ grows, increases the 
validity of the large field form (\ref{Vlead}), and makes the classical contribution 
ever less significant.

\begin{figure}[H]
\includegraphics[width=6cm,height=6cm]{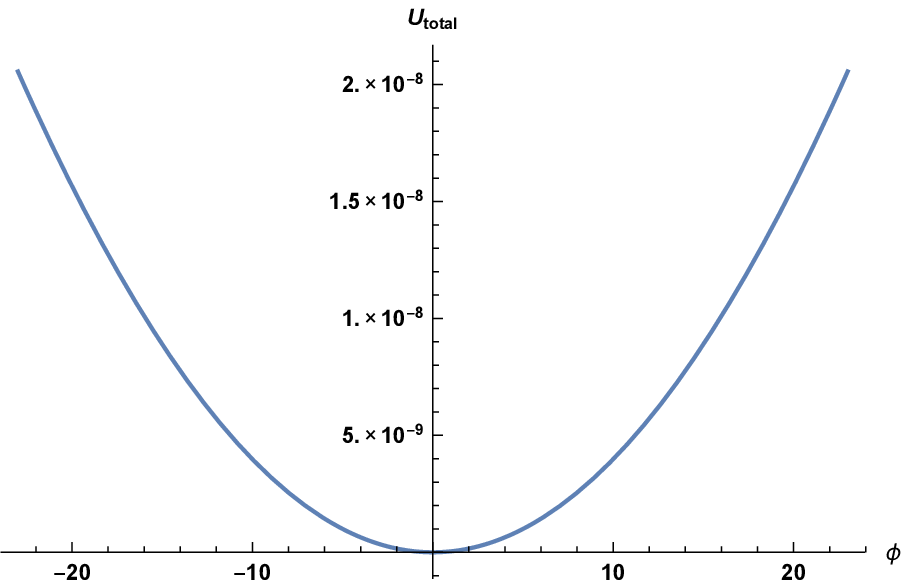}
\hspace{0.3cm}
\includegraphics[width=6cm,height=6cm]{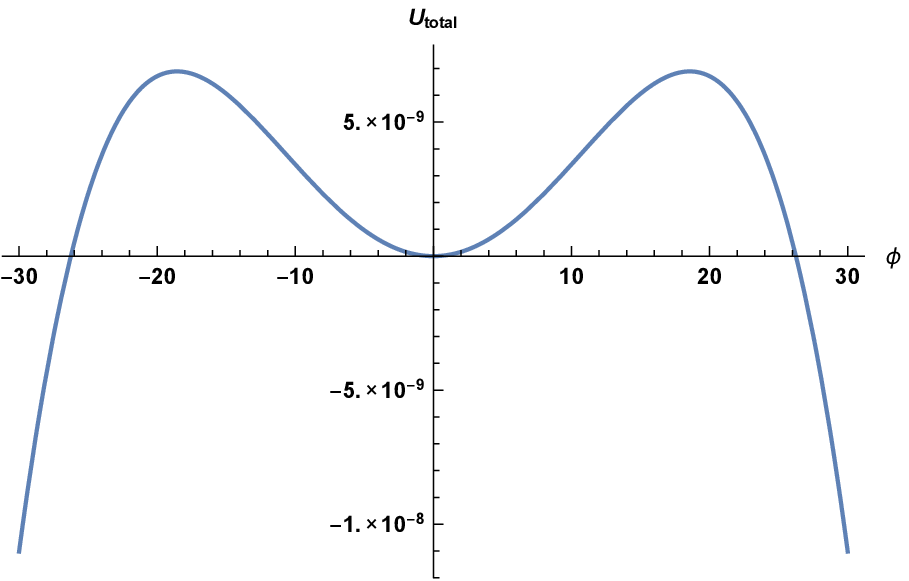}
\caption{The left hand figure shows $U(\phi) + U_{\rm eff}(\phi,\chi_0)$ for the
small coupling of $q^2 = 5 \times 10^{-7} < q^2_{*}(\phi_0)$. The right hand graph 
shows the result of making the coupling slightly larger $q^2 = 2.6 \times 10^{-6}
> q^2_{*}(\phi_0)$.} 
\label{potentials}
\end{figure}

Figure~\ref{potentials} compares the initial shapes of the total potential
$U(\phi) + U_{\rm eff}(\phi,\chi)$ for $q^2 = 5 \times 10^{-7} < q^2_{*}(\phi_0)$ 
and for $q^2 = 2.6 \times 10^{-6} > q^2_{*}(\phi_0)$. 
We numerically checked that the large field estimate (\ref{qcrit}) of
$q^2_{*}(\phi_0) \simeq 2.4 \times 10^{-6}$ in fact marks the crossover point 
between the two regimes. Note that, whereas the large field limiting form 
(\ref{Vlead}) is independent of the classical potential, the threshold value
of the coupling constant $q^2_{*}(\phi_0)$ can be very different for different
classical models. In particular, the very flat potentials favored by current 
data \cite{Aghanim:2018eyx} correspond to much smaller values of $q^2_{*}(\phi_0)$.

\section{\large Conclusions}

Cosmological Coleman-Weinberg potentials are the price scalar-driven inflation
pays for efficiently communicating the kinetic energy of the inflaton to ordinary
matter. Although they take the same $\pm \varphi^4 \ln(\varphi)$ form as their
famous flat space antecedents \cite{Coleman:1973jx}, explicit results on de Sitter
background take the form of $H^4$ times very complicated functions of $\varphi/H$
\cite{Candelas:1975du,Allen:1983dg}. A recent computation \cite{Kyriazis:2019xgj} 
strongly supports the idea that de Sitter results remain approximately valid when 
the constant Hubble parameter of de Sitter is replaced by the evolving $H(t)$ of
realistic inflation. The conundrum for scalar-driven inflation is that cosmological
Coleman-Weinberg potentials are too large and too steep for successful inflation
\cite{Green:2007gs}, while the degree to which they can be subtracted off using 
local, lower derivative counterterms is limited by their dependence on the Hubble 
parameter \cite{Miao:2015oba}. Unacceptable results follow from subtractions 
involving just the inflaton \cite{Liao:2018sci}, or the inflaton and the Ricci 
scalar \cite{Miao:2019bnq}. This paper has been devoted to a different approach in 
which no subtractions are made but one attempts instead to cancel the positive 
potentials induced by bosons with the negative potentials induced by fermions.

In section 2.1 we reviewed the effective potential (\ref{Veff}-\ref{bdef}) induced 
for a model (\ref{totalL}) in which a complex scalar inflaton is coupled to fermions 
(with Yukawa constant $h^2$) and to vector bosons (with charge $q$). Choosing
$h^2 = \sqrt{3} \, q^2$ makes the positive large field form induced by vector bosons 
cancel the negative large field form induced by fermions. Unfortunately, the residual
(\ref{Ueff}) is negative and still large enough to overwhelm classical inflation
and make the universe suffer a Big Rip singularity unless the coupling constant $q^2$ 
is chosen smaller than the minuscule value of $q^2_{*}(\phi_0) \simeq 2.4 \times 
10^{-6}$. Figures \ref{smallq} and \ref{largeq} show what happens for couplings below 
and above this critical value.

The ultimate problem with our cancellation scheme is that the cosmological 
Coleman-Weinberg potentials induced by different sorts of particles are all slightly 
different on the background of an expanding universe, even though they have the same 
functional form (up to a sign) on flat space background. Despite the unsatisfactory 
nature of our results so far we believe there is hope for a better outcome from two
modifications. The first is to make a different choice for the finite part of the 
quartic counterterm (\ref{deltalambda}) so as to cancel the $\varphi^4$ term 
(\ref{Vlead}) of the residual result. In that case the leading large field behavior
would go like $H^2 \varphi^2 \ln(\vert\varphi\vert)$, and the positive bosonic 
contribution should dominate the negative fermionic contribution. This residual 
should still be larger than the classical potential unless $q^2$ is very small, so 
one would need to worry about unacceptable large contributions to the power spectrum.
Another concern is that the original counterterms (\ref{deltaxi}-\ref{deltalambda})
were chosen to keep the small field expansions of the effective potential weak so
as not to disrupt late time physics \cite{Miao:2015oba}.

An even more hopeful modification is to involve scalars in the cancellation scheme.
In addition to its coupling $c^2$ with the inflaton, a real scalar $\Phi$ possess an
additional parameter $\Delta \xi$ characterizing its coupling to the Ricci scalar,
\begin{equation}
\mathcal{L}_{\rm scalar} = -\frac12 \partial_{\mu} \Phi \partial_{\nu} \Phi g^{\mu\nu}
\sqrt{-g} - \frac12 \Bigl(1 \!+\! \Delta \xi\Bigr) \Phi^2 R \sqrt{-g} - \frac14 c^2
\Phi^2 \vert\varphi\vert^2 \sqrt{-g} \; . \label{scalarL}
\end{equation}
We can use this extra parameter to vary the results. If the renormalization constants
are chosen according to the same scheme \cite{Miao:2015oba} that was used for fermions 
and vector bosons the resulting effective potential takes the form,
\begin{equation}
\Delta V_{\rm eff} = \frac{H^4}{64 \pi^2} \Biggl\{ s(c^2 z^2) + \Bigl[2 \Delta \xi
c^2 z + \frac{c^4 z^2}{4} \Bigr] \ln\Bigl( \frac{H^2}{\mu^2}\Bigr) \Biggr\} \; ,
\label{DeltaVeff}
\end{equation}
where $z \equiv \vert \varphi\vert^2/H^2$. The function $s(y)$ is \cite{Miao:2015oba},
\begin{eqnarray}
\lefteqn{s(y) = -\Bigl[ \psi(\nu_{+}) \!+\! \psi(\nu_{-})\Bigr] \Bigl[2 \Delta \xi y
\!+\! \frac14 y^2\Bigr] + \Bigl[ \psi'(\nu_{+}) - \psi'(\nu_{-})\Bigr] 
\frac{\frac12 \Delta \xi y^2}{\sqrt{1 \!-\! 8 \Delta \xi}} } \nonumber \\
& & \hspace{0cm} + \int_{0}^{y} \!\! dx \, \Bigl(2 \Delta \xi \!+\! \frac{x}{2}\Bigr)
\Biggl[ \psi\Biggl( \frac12 + \sqrt{\frac14 \!-\! 2 \Delta \xi \!-\! \frac{x}{2} }
\, \Biggr) + \psi\Biggl( \frac12  \sqrt{\frac14 \!-\! 2 \Delta \xi \!-\! \frac{x}{2} }
\, \Biggr) \Biggr] . \qquad \label{sdef}
\end{eqnarray}
By employing a number of scalars with different values of $c^2$ and $\Delta \xi$ it 
may be possible to cancel enough terms in the large field expansion to make the 
residual harmless for at least some models of inflation.

Before closing we should comment on the specious argument sometimes advanced to
dismiss the possibility that cosmological Coleman-Weinberg potentials pose any 
danger for inflation. The argument begins by observing that the inflaton is large 
during the early stages of inflation, endowing the ordinary matter to which it is 
coupled with a large mass. Of course that is correct, but the second part of the 
argument mistakenly concludes that such a large mass suppresses quantum fluctuations 
of ordinary matter so that they cannot make significant corrections to the effective 
potential. This reasoning is belied by the classic flat space results which are 
known to go like $\pm \varphi^4 \ln(\varphi)$ \cite{Coleman:1973jx}. In fact the 
argument reveals a confusion about what the effective potential represents. At one 
loop order the primitive, unrenormalized, effective potential is just the integral 
of the 0-point energies from each plane wave mode $\vec{k}$,
\begin{equation}
V_{\rm prim} = \int \!\! \frac{d^3k}{(2\pi)^3} \, \frac12 \hbar \omega(\vec{k}) \; .
\end{equation}
In flat space $\omega^2 = M^2 + \Vert \vec{k}\Vert^2$, so increasing the mass $M$
must increase, not decrease, the effective potential. This is evident if we perform
the flat space computation with a momentum cutoff,
\begin{eqnarray}
\lefteqn{\Bigl( V_{\rm prim}\Bigr)_{\rm flat} = \frac{4\pi}{8 \pi^3} \!\! 
\int_{0}^{\Lambda} \!\! dk \, k^2 \sqrt{k^2 +M^2} \; , } \\
& & \hspace{1cm} = \frac1{16 \pi^2} \Biggl\{(2 \Lambda^3 + \Lambda M^2) 
\sqrt{\Lambda^2 + M^2} - M^4 \ln\Biggl[\frac{\Lambda + \sqrt{\Lambda^2 + M^2}}{M} 
\, \Biggr] \Biggr\} . \label{prim} \qquad
\end{eqnarray}
The primitive contribution (\ref{prim}) is renormalized by a counterterm with
scale $\mu$,
\begin{equation}
\Bigl( \Delta V\Bigr)_{\rm flat} = -\frac1{16 \pi^2} \Biggl\{ 2 \Lambda^4 + 2 M^2
\Lambda^2 + \frac14 M^4 - M^4 \ln\Bigl( \frac{2 \Lambda}{\mu}\Bigr) \Biggr\} .
\label{counter}
\end{equation}
Adding the counterterm (\ref{counter}) and taking the unregulated limit gives
the familiar result \cite{Coleman:1973jx},
\begin{equation}
\lim_{\Lambda \rightarrow \infty} \Bigl( V_{\rm prim} + \Delta V\Bigr)_{\rm flat}
= \frac1{16 \pi^2} M^4 \ln\Bigl( \frac{M}{\mu}\Bigr) \; . \label{flatCWP}
\end{equation}
From the integral (\ref{prim}) we can even see that the largest part of the final
answer (\ref{flatCWP}) derives from momenta up to $k \sim M$, so large values of
$M$ result in high momentum modes contributing. The expansion of the universe 
changes the analysis in important ways, but it does not alter the basic fact that 
increasing the inflaton field strength, which increases $M$, increases the effective
potential that is induced. Note also that increasing $\omega$ does suppress the 
length of time virtual particles can persist, so it really does suppress many 
other quantum effects, just not the effective potential.

\vskip 1cm

\centerline{\bf Acknowledgements}

This work was partially supported by Taiwan MOST grants 
108-2112-M-006-004 and 107-2119-M-006-014; by NSF grants PHY-1806218 
and PHY-1912484; and by the Institute for Fundamental Theory at the 
University of Florida.

\section{Appendix: \large Hubble Proxies}

Although loop contributions to the effective action are not generally local,
any subtraction of them must be local, generally coordinate invariant, and 
free of the instabilities that plague interacting continuum field theories 
which contain nondegenerate higher time derivatives \cite{Woodard:2006nt}.
In short, any subtraction must be acceptable as part of the classical action.
One cannot otherwise compute inflationary perturbations, and couple inflation
to ordinary matter.

The Hubble parameter which characterizes cosmological expansion is not a local,
generally coordinate invariant functional of the metric. Any subtraction scheme
that attempts to exactly cancel cosmological Coleman-Weinberg potentials must
exploit a proxy for the Hubble parameter, that is, an invariant functional of 
fields which reduces to $H(t)$ in the homogeneous and isotropic geometry 
(\ref{background}) of an expanding universe. The usual technique for constructing
such a proxy is based on taking the divergence of a normalized, timelike vector 
field $u^{\mu}(x)$ \cite{Geshnizjani:2002wp},
\begin{equation}
g_{\mu\nu}(x) u^{\mu}(x) u^{\nu}(x) = -1 \qquad \Longrightarrow \qquad
\mathcal{H}(x) \equiv -\frac13 D_{\mu} u^{\mu}(x) \; . \label{utoH}
\end{equation}
The question then becomes, where are we to find the normalized, timelike vector
field $u^{\mu}(x)$? The literature provides three sorts of answers:
\begin{enumerate}
\item{One might construct it from the gradient of the inflaton and the metric 
\cite{Geshnizjani:2002wp},
\begin{equation}
u^{\mu}(x) \equiv -\frac{g^{\mu\nu}(x) \partial_{\nu} \varphi(x)}{\sqrt{-
g^{\alpha\beta}(x) \partial_{\alpha} \varphi(x) \partial_{\beta} \varphi(x)}} \; ;
\label{ufromphi}
\end{equation}}
\item{One might construct it purely from the metric by making the replacement 
$\varphi(x) \rightarrow \Phi[g](x)$ in the previous construction (\ref{ufromphi}),
where $\Phi[g](x)$ is defined as the solution of a suitable differential equation 
with boundary conditions specified on some initial value surface 
\cite{Miao:2017vly}; or}
\item{One might introduce $u^{\mu}(x)$ as a new fundamental field in an
Einstein-Aether theory \cite{Jacobson:2000xp}.}
\end{enumerate}

Each of the three Hubble proxies has problems associated with the fact that it 
is fundamentally some functional of fields which only reduces to the Hubble
parameter after the equations of motion are used and the geometry is specialized 
to (\ref{background}). Hence the variations that give the field equations (which 
must be taken {\it before} specialization) can introduce undesirable effects. For 
example, constructing the timelike vector from the gradient of the inflaton 
obviously involves higher time derivatives (and hence Ostrogradsky instabilities) 
when (\ref{ufromphi}) is substituted in (\ref{utoH}). The 2nd construction is 
unacceptable because the Hubble proxy is not even a local functional of the 
metric. The 3rd construction will see variations of the Hubble proxy making 
extensive changes in the $u^{\mu}(x)$ field equation. Even if these changes 
should prove benign, it is known that cosmological Coleman-Weinberg potentials 
depend in a complicated way on the first slow roll parameter $\epsilon(t) = 
-\dot{H}/H^2$, in addition to just the Hubble parameter \cite{Kyriazis:2019xgj}. 
Attempting to include this dependence will introduce higher time derivatives of 
$u^{\mu}(x)$, which must engender Ostrogradsky instabilities.


\begin{thebibliography}{99}

\bibitem{Geshnizjani:2011dk}
G.~Geshnizjani, W.~H.~Kinney and A.~Moradinezhad Dizgah,
JCAP \textbf{11} (2011), 049
doi:10.1088/1475-7516/2011/11/049
[arXiv:1107.1241 [astro-ph.CO]].

\bibitem{Kinney:2012zbd} 
  W.~H.~Kinney, G.~Geshnizjani and A.~Moradinezhad Dizgah,
  ``Inflation, Or What?,'' in {\it 47th Rencontres de Moriond on Cosmology:
  La Thuile, Italy, March 10-17, 2012}, ed. E. Aug\'e, J. Dumarchez and J. T. T. Van,
  pp. 179-184.
  
\bibitem{Akrami:2018odb} 
  Y.~Akrami {\it et al.} [Planck Collaboration],
  arXiv:1807.06211 [astro-ph.CO].

\bibitem{Ijjas:2013vea} 
  A.~Ijjas, P.~J.~Steinhardt and A.~Loeb,
  Phys.\ Lett.\ B {\bf 723}, 261 (2013)
  doi:10.1016/j.physletb.2013.05.023
  [arXiv:1304.2785 [astro-ph.CO]].
      
\bibitem{Guth:2013sya} 
  A.~H.~Guth, D.~I.~Kaiser and Y.~Nomura,
  Phys.\ Lett.\ B {\bf 733}, 112 (2014)
  doi:10.1016/j.physletb.2014.03.020
  [arXiv:1312.7619 [astro-ph.CO]].

\bibitem{Linde:2014nna} 
  A.~Linde,
  doi:10.1093/acprof:oso/9780198728856.003.0006
  arXiv:1402.0526 [hep-th].

\bibitem{rebuttal}
  A.~H.~Guth {\it et al.} [33 co-authors], ``A Cosmic Controversy,''
  (Letter to the Editor), {\it Scientific American} (May 10, 2017).
  
\bibitem{Ijjas:2014nta} 
  A.~Ijjas, P.~J.~Steinhardt and A.~Loeb,
  Phys.\ Lett.\ B {\bf 736}, 142 (2014)
  doi:10.1016/j.physletb.2014.07.012
  [arXiv:1402.6980 [astro-ph.CO]].
  
\bibitem{Coleman:1973jx} 
  S.~R.~Coleman and E.~J.~Weinberg,
  Phys.\ Rev.\ D {\bf 7}, 1888 (1973).
  doi:10.1103/PhysRevD.7.1888
  
\bibitem{Green:2007gs} 
  D.~R.~Green,
  Phys.\ Rev.\ D {\bf 76}, 103504 (2007)
  doi:10.1103/PhysRevD.76.103504
  [arXiv:0707.3832 [hep-th]].

\bibitem{Miao:2015oba} 
  S.~P.~Miao and R.~P.~Woodard,
  JCAP {\bf 1509}, 022 (2015)
  doi:10.1088/1475-7516/2015/09/022, 10.1088/1475-7516/2015/9/022
  [arXiv:1506.07306 [astro-ph.CO]].

\bibitem{Kyriazis:2019xgj} 
  A.~Kyriazis, S.~P.~Miao, N.~C.~Tsamis and R.~P.~Woodard,
  arXiv:1908.03814 [gr-qc].
  
\bibitem{Liao:2018sci} 
  J.~H.~Liao, S.~P.~Miao and R.~P.~Woodard,
  Phys.\ Rev.\ D {\bf 99}, no. 10, 103522 (2019)
  doi:10.1103/PhysRevD.99.103522
  [arXiv:1806.02533 [gr-qc]].

\bibitem{Miao:2019bnq} 
  S.~P.~Miao, S.~Park and R.~P.~Woodard,
  Phys.\ Rev.\ D {\bf 100}, no. 10, 103503 (2019)
  doi:10.1103/PhysRevD.100.103503
  [arXiv:1908.05558 [gr-qc]].

\bibitem{Woodard:2006nt} 
  R.~P.~Woodard,
  Lect.\ Notes Phys.\  {\bf 720}, 403 (2007)
  doi:10.1007/978-3-540-71013-4\_14
  [astro-ph/0601672].

\bibitem{Candelas:1975du} 
  P.~Candelas and D.~J.~Raine,
  Phys.\ Rev.\ D {\bf 12}, 965 (1975).
  doi:10.1103/PhysRevD.12.965

\bibitem{Allen:1983dg} 
  B.~Allen,
  Nucl.\ Phys.\ B {\bf 226}, 228 (1983).
  doi:10.1016/0550-3213(83)90470-4

\bibitem{Miao:2006pn} 
  S.~P.~Miao and R.~P.~Woodard,
  Phys.\ Rev.\ D {\bf 74}, 044019 (2006)
  doi:10.1103/PhysRevD.74.044019
  [gr-qc/0602110].

\bibitem{Prokopec:2007ak} 
  T.~Prokopec, N.~C.~Tsamis and R.~P.~Woodard,
  Annals Phys.\  {\bf 323}, 1324 (2008)
  doi:10.1016/j.aop.2007.08.008
  [arXiv:0707.0847 [gr-qc]].
  
\bibitem{Simon:1990ic} 
  J.~Z.~Simon,
  Phys.\ Rev.\ D {\bf 41}, 3720 (1990).
  doi:10.1103/PhysRevD.41.3720

\bibitem{Aghanim:2018eyx} 
  N.~Aghanim {\it et al.} [Planck Collaboration],
  arXiv:1807.06209 [astro-ph.CO].

\bibitem{Geshnizjani:2002wp}
G.~Geshnizjani and R.~Brandenberger,
Phys. Rev. D \textbf{66}, 123507 (2002)
doi:10.1103/PhysRevD.66.123507
[arXiv:gr-qc/0204074 [gr-qc]].

\bibitem{Miao:2017vly}
S.~Miao, N.~Tsamis and R.~Woodard,
Phys. Rev. D \textbf{95}, no.12, 125008 (2017)
doi:10.1103/PhysRevD.95.125008
[arXiv:1702.05694 [gr-qc]].

\bibitem{Jacobson:2000xp}
T.~Jacobson and D.~Mattingly,
Phys. Rev. D \textbf{64}, 024028 (2001)
doi:10.1103/PhysRevD.64.024028
[arXiv:gr-qc/0007031 [gr-qc]].

\end{thebibliography}
\end{document}